\documentstyle[12pt]{article}
\begin{document}

\title{Radiative Fermion Mass Hierarchy in a Non-supersymmetric Unified Theory}

\author{{\bf S.M. Barr} \\
Bartol Research Institute \\ University of Delaware \\
Newark, Delaware 19716}

\date{\today}

\maketitle

\begin{abstract}
In non-supersymmetric grand unified models a ``radiative
fermion mass hierarchy" can be achieved in which the spectrum of quark and lepton
masses is determined entirely by physics at the unification scale, with many
relations following from the unified gauge symmetry, and with the masses of
the lightest family arising from loops. A simple, realistic,
and predictive model of this kind is presented. A ``doubly lopsided" structure, known
to lead to bilarge neutrino mixing, plays a crucial role in the radiative hierarchy.
\end{abstract}

\newpage

The masses of the charged quarks and leptons span a large range, the
electron mass being only $2.9 \times 10^{-6}$ of the $t$ quark mass. The masses
of fermions
of the same type in successive families typically differ by about two orders
of magnitude. These facts have suggested to many people the possibility of a
``radiative hierarchy",
i.e. a hierarchy in which the heavier fermions get mass from tree diagrams while the
lighter fermions get mass from loop diagrams \cite{rad}. Indeed, this speculation is very old,
going back to the observation that $m_e/m_{\mu} = O(\alpha)$.

In the early 1980s, several papers \cite{radBarr} showed that the idea of a radiative fermion mass
hierarchy can be implemented in grand unified theories (GUTS) in such a way that the
structure of the observed quark and lepton masses is entirely determined by physics
at the unification scale $M_{GUT}$. In these models the loop diagrams that generate
the radiative fermion masses contain virtual particles with masses of order
$M_{GUT}$. Interest in such models waned with the growing popularity
of the idea of low-energy supersymmetry. In supersymmetric models, radiative
corrections to the quark and lepton mass matrices would be highly suppressed by the
non-renormalization theorem.

There are a number of reasons why one should still take seriously the
idea of nonsupersymmetric grand unified models. The main one, of course, is that
low-energy supersymmetry has not yet been discovered. Moreover, models with
low-energy supersymmetry face several well-known challenges, such as the
``$\mu$ problem", the ``SUSY flavor problem", and the ``SUSY CP problem".
While supersymmetry does lead to an impressive unification of gauge couplings
and thus seems to go hand-in-hand with the idea of unification,
supersymmetric grand unified theories are not without their own difficulties,
notably the danger of excessive proton decay mediated by the exchange of Higgsinos.
And the main theoretical problem solved by low-energy supersymmetry, namely
the tuning of the Higgs mass, may not be so dreadful in the context of
the ``landscape", where it may be explained ``anthropically" \cite{abds}.

Another reason that interest in radiative hierarchy schemes waned in
the 1980s is that they seemed to require many extra fermions to play
the role of virtual particles in loops, and thus it seemed that the
models must be rather complicated and unpredictive.  For example, in
a conference talk in 1982, H. Georgi remarked, ``A more serious
problem with models of this type is that it is hard to find any
without inventing an enormous number of superheavy fermions. I like
the abstract idea much better than any of the realizations I have
found." \cite{georgi} In this letter a relatively simple
non-supersymmetric grand unified model is proposed that leads to a
radiative fermion mass hierarchy. Very few superheavy fermions are
required, in fact only a family-antifamily pair and a vector
multiplet. In this model virtually every qualitative feature of the
pattern of quark and lepton masses and mixings is reproduced from a
rather simple underlying structure. The interfamily mass ratios and
mixing angles, though spanning a large range, are accounted for
without the ad hoc introduction of small parameters, since the
``loop factors" of $1/16 \pi^2$ explain much of the hierarchical
structure.

The basic structure of the model discussed here was actually proposed
in an earlier paper \cite{bb2002}. However, in
that paper it was assumed that all the effective fermion mass operators arise at
tree level. No explanation was therefore given why certain elements of
the mass matrices were of order $10^{-2}$ compared to the largest elements. Moreover,
the existence of the small elements could only be accounted for as tree-level
effects by the
introduction of a significant amount of additional structure in the Yukawa
sector of the theory. Here it is shown that those small elements
arise very naturally as radiative corrections, without such additional
structure. Thus a simpler model leads to the same results, but with the
bonus that both the existence of the small elements and their size is naturally
explained.

The model is an $SO(10)$ grand unified theory in which the quarks and
leptons are contained in the following (left-handed) multiplets:
${\bf 16}_{i = 1,2,3} + \;\;
({\bf 16} + \overline{{\bf 16}} + {\bf 10})$. The ``extra" fermion multiplets in
the parentheses form real representations of $SO(10)$, and so the low-energy
spectrum consists of just three chiral families, as observed. The Dirac mass matrices
of the up-type quarks, down-type quarks, charged leptons, and neutrinos (denoted by
$M_U$, $M_D$, $M_L$, and $M_N$, respectively) arise from the following set of Yukawa
terms:

\begin{equation}
\begin{array}{ccl}
{\cal L}_{Yuk} & = & M( \overline{{\bf 16}} \; {\bf 16}) + M_{10} ( {\bf 10} \; {\bf 10}).
\\ & & \\
& + & a( \overline{{\bf 16}} \; {\bf 16}_3 ) {\bf 45}_H + \sum_{i=1,2} c_i ( {\bf 10} \;
{\bf 16}_i) {\bf 16}_{iH} \\ & & \\
& + & h_{33} ( {\bf 16}_3 {\bf 16}_3 ) {\bf 10}_H + h_2 ( {\bf 16} \; {\bf 16}_2 )
{\bf 10}_H + h_3 ( {\bf 10} \; {\bf 16}_3)
{\bf 16}'_H \\ & & \\
& + & h ( {\bf 16} \; {\bf 16} ) {\bf 10}'_H. \end{array}
\end{equation}

\noindent
The terms on the first line of Eq. (1) are the $O(M_{GUT})$ masses of the extra
fermion multiplets; the terms on the second line contribute $O(M_{GUT})$
masses that mix those extra fermions with the three chiral families ${\bf 16}_i$;
the terms on the third line generate the weak-scale $SU(2)_L \times
U(1)_Y$-breaking masses; and the last term is needed to give radiative masses to
the first family. Higgs multiplets are denoted by the subscript $H$. The Higgs fields
${\bf 16}_{iH}$ obtain vacuum expectation values (VEV) in the ${\bf 1} ({\bf 16})$
direction. (The expression ${\bf p}({\bf q})$ stands for a ${\bf p}$ multiplet of
$SU(5)$ contained in a ${\bf q}$ multiplet of $SO(10)$.) The adjoint
Higgs field ${\bf 45}_H$ obtains a VEV that is proportional to the $SO(10)$
generator $B-L$ (i.e. baryon number minus lepton number).
These two types of Higgs fields would be enough to break $SO(10)$
down to the Standard Model group $SU(3)_c \times SU(2)_L \times U(1)_Y$; however,
there may be other Higgs fields that also contribute to that breaking.

The electroweak gauge symmetry
$SU(2)_L \times U(1)_Y$ is spontaneously broken by the Higgs
multiplets denoted ${\bf 10}_H$, ${\bf 10}'_H$, and ${\bf 16}'_H$ in Eq. (1),
and, more specifically, by the neutral components of the $Y/2= -1/2$ doublets contained
in $\overline{{\bf 5}}({\bf 10}_H)$, $\overline{{\bf 5}}({\bf 10}'_H)$, and
$\overline{{\bf 5}}({\bf 16}'_H)$, and the neutral components of the $Y/2 = +1/2$
doublets contained in ${\bf 5}({\bf 10}_H)$ and ${\bf 5}({\bf 10}'_H)$. Of course,
in the low-energy effective theory, which is just the Standard Model, there is
only one Higgs doublet, which is some linear combination of these doublets (and
their hermitian conjugates).

There are several layers to the quark and lepton mass matrices that arise
from Eq. (1). The first layer comes simply from the term $h_{33} ( {\bf 16}_3
\; {\bf 16}_3 ) {\bf 10}_H$ and gives contributions of the form

\begin{equation}
\begin{array}{l}
M_U^{(1)} = M_N^{(1)} = \left( \begin{array}{ccc}
0 & 0 & 0 \\ 0 & 0 & 0 \\ 0 & 0 & 1 \end{array} \right) m_U, \\ \\
M_D^{(1)} = M_L^{(1)} = \left( \begin{array}{ccc}
0 & 0 & 0 \\ 0 & 0 & 0 \\ 0 & 0 & 1 \end{array} \right) m_D, \end{array}
\end{equation}

\noindent
where $m_U \equiv h_{33} \langle {\bf 5}({\bf 10}_H) \rangle$ and
$m_D \equiv h_{33} \langle \overline{{\bf 5}}({\bf 10}_H) \rangle$.

The second layer comes from integrating out the family-antifamily pair
$\overline{{\bf 16}} + {\bf 16}$. The antifamily $\overline{{\bf 16}}$
appears in two mass terms from Eq. (1),
which can be combined as follows: $\overline{{\bf 16}} (M {\bf 16} +
a \langle {\bf 45}_H \rangle {\bf 16}_3)$. These terms have the effect of
mixing the ${\bf 16}$ with the ${\bf 16}_3$. One family
obtains an $O(M_{GUT})$ mass, while the other (denoted by the index $3'$)
remains light. (From now on, primed indices will
be used to denote the light families that remain after the superheavy
fermions have been
integrated out.) Thus, the ${\bf 16}$ with no index
has some of the third light family mixed in with it. The amount of this mixing
is proportional to $B-L$, since $\langle {\bf 45}_H \rangle \propto B-L$,
and it is thus three times stronger for the
leptons than the quarks. As a result, the term $h_2 ( {\bf 16} \; {\bf 16}_2)
{\bf 10}_H$ from Eq. (1) leads to an effective operator of the form
$({\bf 16}_{3'} {\bf 16}_{2'}) {\bf 10}_H {\bf 45}_H/M_{GUT}$, which
in turn produces contributions to the effective low-energy mass matrices of the
form

\begin{equation}
M_{U/D}^{(2)} = \left( \begin{array}{ccc} 0 & 0 & 0 \\
0 & 0 & \epsilon/3 \\ 0 & - \epsilon/3 & 0 \end{array}
\right) m_{U/D}, \;\;
M_{N/L}^{(2)} = \left( \begin{array}{ccc} 0 & 0 & 0 \\
0 & 0 & - \epsilon \\ 0 & \epsilon & 0 \end{array}
\right) m_{U/D}.
\end{equation}

\noindent
The parameter $\epsilon$ is ``naturally" of order one (or slightly smaller, since
it is proportional to the sine of the mixing angle between ${\bf 16}$
and ${\bf 16}_3$). To fit the observed masses its actual value must be about
0.15.

In a similar way, effective operators arise from integrating out the
$SO(10)$-vector multiplet of quarks and leptons, ${\bf 10}$. This
multiplet contains a $\overline{{\bf 5}} + {\bf 5}$ of $SU(5)$. The ${\bf 5}({\bf
10})$ appears in several mass terms from Eq. (1), which can be
combined as ${\bf 5}({\bf 10}) [ M_{10} \overline{{\bf 5}} ({\bf
10}) + \sum_{i=1,2} c_i \langle {\bf 1}({\bf 16}_{iH}) \rangle \overline{{\bf
5}} ({\bf 16}_i) ]$. These terms have the effect of mixing the
$\overline{{\bf 5}}({\bf 10})$ with the $\overline{{\bf 5}} ({\bf
16}_1)$ and $\overline{{\bf 5}}({\bf 16}_2)$. One linear combination
of them obtains an $O(M_{GUT})$ mass, while the other two linear combinations
are in the
light families and denoted $\overline{{\bf 5}}_{1'}$ and
$\overline{{\bf 5}}_{2'}$. Consequently, the $SO(10)$-vector of fermions
(${\bf 10}$) has mixed in with it some of the first two families of
right-handed down quarks and left-handed leptons (these being the
particle types contained in a $\overline{{\bf 5}}$ of $SU(5)$). That
means that the term $h_3 ({\bf 10} \; {\bf 16}_3 ) {\bf 16}'_H$ in
Eq. (1) leads to effective mass terms of the form $(C_1
\overline{{\bf 5}}_{1'} + C_2 \overline{{\bf 5}}_{2'}) {\bf 10}_{3'} m_D$.
This gives

\begin{equation}
M_N^{(3)} = M_U^{(3)} = 0, \;\;\; M_D^{(3)} = M_L^{(3)T} = \left(
\begin{array}{ccc} 0 & 0 & 0 \\ 0 & 0 & 0 \\ C_1 & C_2 & 0
\end{array} \right) m_D.
\end{equation}

\noindent
Note the crucial fact that only $M_D$ and $M_L$ get these contributions.
The parameters $C_1$ and $C_2$ are naturally of order one. The actual
fit to the data require $C_1 \simeq 0.8$ and $C_2 \simeq 1.2$.

The full tree-level mass matrices, which are obtained by adding the
three layers given in Eqs. (2)-(4), have the form

\begin{equation}
\begin{array}{ll}
M_U = \left( \begin{array}{ccc} 0 & 0 & 0 \\
0 & 0 & \epsilon/3 \\ 0 & - \epsilon/3 & 1 \end{array}
\right) \; m_U, \;\;\; & M_D = \left( \begin{array}{ccc}
0 & 0 & 0 \\ 0 & 0 & \epsilon/3 \\
C_1 & C_2 - \frac{\epsilon}{3} & 1 \end{array} \right) \; m_D, \\ & \\
M_N = \left( \begin{array}{ccc} 0 & 0 & 0 \\
0 & 0 & - \epsilon \\ 0 & \epsilon & 1 \end{array}
\right) \; m_U, \;\;\; & M_L = \left( \begin{array}{ccc}
0 & 0 & C_1 \\ 0 & 0 & C_2 - \epsilon \\
0 & \epsilon & 1 \end{array} \right) \; m_D. \end{array}
\end{equation}

\noindent
The convention here is that the mass matrices are multiplied from the left by
the left-handed fermions and from the right by the right-handed fermions.
These equations for the quark and lepton mass matrices are approximate.
The exact expressions involve factors, such as
$1/\sqrt{1 + (a \langle {\bf 45}_H \rangle/M)^2}$ and
$1/\sqrt{1 + (\sum_i c_i \langle {\bf 16}'_H \rangle/M_{10})^2}$, which
are essentially just the cosines of angles describing
the mixing between the extra fermions ${\bf 16} + \overline{{\bf 16}} + {\bf 10}$
and the three chiral families ${\bf 16}_i$. If one assumes these mixing angles
are small, these cosine factors may be set to one. (The fact that $\epsilon \simeq 0.15$
would most simply be explained by saying that $a \langle {\bf 45}_H \rangle/M$
is of that order.)

Before turning to the radiative effects that generate the masses and mixings
of the first family, let us see how the simple tree-level structure of
Eq. (5) reproduces
the known pattern of masses and mixings of the second and third families.
The first thing to notice is that in the limit of $\epsilon \rightarrow 0$
only one family has mass. The mass of the $t$ quark is just $m_U$, whereas the
$b$ quark and $\tau$ lepton both have mass $\sqrt{ 1 + |C_1|^2 + |C_2|^2} \; m_D$.
The well-known successful $SU(5)$ relation $m_b^0 \cong m_{\tau}^0$ is thus obtained in
this model. The superscipt $0$ refers to relations at the scale $M_{GUT}$.
The large off-diagonal elements $C_1$ and $C_2$ that appear
asymmetrically in $M_D$ and $M_L$ are the characteristic features of ``lopsided"
models \cite{lopsided}. Having both large $C_1$ and $C_2$ makes this a ``doubly lopsided"
model. Such a structure, as is well-known \cite{bb2002}, leads to the ``bi-large" pattern of
neutrino mixing that has been observed. This happens by the diagonalization of
the {\it charged} lepton mass matrix $M_L$. In diagonalizing $M_L$, the large
element $C_1$ must be eliminated by a rotation from the left by an angle
$\theta_{sol}$ in the 12 plane, where $\tan \theta_{sol} = -C_1/C_2$. In the process,
$C_2$ is replaced by $C \equiv \sqrt{|C_1|^2 + |C_2|^2}$. This element must then be
eliminated by a rotation acting on $M_L$ from the left by an angle $\theta_{atm}$
in the 23 plane, where $\tan \theta_{atm} = - C$. The net result, if no other
rotations of leptons were done, would be a neutrino mixing matrix of the form

\begin{equation}
U_{MNS} = \left( \begin{array}{ccc}
c_s & s_s & 0 \\
- c_a s_s & c_a c_s & s_a \\ s_a s_s & -s_a c_s & c_a \end{array}
\right),
\end{equation}

\noindent
where $s_a \equiv \sin \theta_{atm}$, $c_a \equiv \cos \theta_{atm}$,
$s_s \equiv \sin \theta_{sol}$, and $c_s \equiv \cos \theta_{sol}$. As can be seen
from the form of $M_N$ in Eq. (5), there are rotations in the
23 plane of $O(\epsilon)$
required to diagonalize the neutrino mass matrix. And there will be
a further rotation of order $\sqrt{m_e/m_{\mu}}$ in the 12 plane required to
complete the diagonalization of the charged lepton mass matrix $M_L$. So the
neutrino mixing matrix will not be of exactly the form shown in Eq. (6). In
particular, the 13 angle will not be exactly zero, but of order $0.1$.

There are no corresponding large contributions to the CKM mixing angles of
the quarks, since the rotations required to eliminate the large
off-diagonal elements $C_1$ and $C_2$ from $M_D$ involve only the
{\it right-handed} quarks. This is, of course, the way that lopsided models
account for the fact that the observed quark mixings are much smaller than the
neutrino mixings. These rotations acting on $M_D$ from the right, bring it to
the form

\begin{equation}
M'_D = \left( \begin{array}{ccc}
0 & 0 & 0 \\ 0 & - s_a \epsilon/3 & c_a \epsilon/3 \\
0 & 0 & 1/c_a \end{array} \right) \; m_D.
\end{equation}

\noindent
We have taken the rotation angles here also to be to be $\theta_{sol}$ and $\theta_{atm}$,
thus neglecting the slight difference between the 23 element of $M_L$
and the 32 element of $M_D$, which is an effect higher order in
$\epsilon$. One sees that a 22 element is induced in $M'_D$, thus
giving the strange quark a mass of
$O(\epsilon)$. In particular, $m_s^0/m_b^0 \cong \sin \theta_{atm} \cos \theta_{atm}
(\epsilon/3) \simeq \epsilon/6$. The corresponding rotations acting
on $M_L$ induce a 22 element that is 3 times larger (because of the
factor of $(B-L)$ that multiplies the parameter $\epsilon$ in these matrices).
This gives the famous Georgi-Jarlskog relation $m_{\mu}^0 \cong 3m_s^0$ \cite{gj}, which is
known to work reasonably well.

Because there are no large off-diagonal elements (like $C_1$, $C_2$)
in the up quark mass matrix $M_U$ there
is no $O(\epsilon)$ 22 element induced in $M_U$. Rather,
$m_c^0/m_t^0 \cong (\epsilon/3)^2$. This accords well with the observed mass
ratios, which do approximately satisfy $m_c/m_t \sim (2m_s/m_b)^2$. The mixing
between the second and third families of quarks, $V_{cb}$, arises from
the mismatch between the 23 rotation required to diagonalize $M_U$, which is
$\epsilon/3$, and the corresponding angle for $M'_D$, which one sees from Eq. (7)
is approximately $\cos^2 \theta_{atm} (\epsilon/3) \sim \epsilon/6$. This gives
$V_{cb} = \sin^2 \theta_{atm} (\epsilon/3) \sim m_s/m_b$, which is qualitatively
correct. However, as will be seen,
there are corrections to $V_{cb}$ coming from the radiative contributions
to $M_D$ that cannot be neglected, so one does not
expect a perfect fit from the tree-level form in Eqs. (5),(7).

The qualitative features of the masses and mixings of the two heavy
families are strikingly reproduced: $\theta_{atm}$, $\theta_{sol}
\sim 1$; $\theta_{13} \ll 1$; $V_{cb} \sim m_s^0/m_b^0 \sim
m_{\mu}/3m_{\tau}^0 \sim \epsilon/6$; and $m_c^0/m_t^0 \sim
\epsilon^2/9$.

In order for the first family of quarks and leptons to obtain mass, there must
be additional non-zero entries in the mass matrices shown in Eq. (5). The
additional entries must make $\det M_D \cong \det M_L \neq 0$.
The reason is that the famous and well-satisfied Georgi-Jarlskog relation $m_e^0/m_{\mu}^0 = \frac{1}{9}
m_d^0/m_s^0$ follows from $m_b^0 = m_{\tau}^0$ and
the other Georgi-Jarlskog relation $m_s^0 = m_{\mu}^0/3$
only if $\det M_D = \det M_L$. Consequently,
a non-zero 11 element of $M_D$ would
require a 11 element of $M_L$ that was $1/3$ as large in order to contribute equally
to $\det M_D$ and $\det M_L$, as can be seen from Eq. (5).
However, that pattern does not arise from any simple
effective
operators. Similarly, a non-zero 12 element of $M_D$ would require a non-zero
21 element of $M_L$ that was $1/3$ as large, which is not easy to achieve for the same
reason. Therefore, given that obviously {\it some} non-zero element
is needed in the first row
of $M_D$ and first column of $M_L$ to give mass to the electron and $d$ quark,
it must be that
$(M_D)_{13} \cong (M_L)_{31} \neq 0$. Such contributions can be arranged
to exist at tree-level by introducing further fermion and scalar multiplets and
appropriate couplings for them, as was done in \cite{bb2002}.
As will now be seen, however, precisely these needed
non-zero elements {\it automatically} arise from
one-gauge-boson-loop diagrams, while such diagrams leave all the other zeros in
the mass matrices unaffected.

The one-gauge-boson-loop diagram in question is shown in Fig. 1(a).
The gauge boson in this diagram is in a ${\bf 10}$ of $SU(5)$ (of
course, it is in the adjoint ${\bf 45}$ of $SO(10)$), so that it
turns ${\bf 10}$'s of $SU(5)$ into $\overline{{\bf 5}}$'s and {\it
vice versa}. That means that it interchanges $d_L \leftrightarrow
d_R$ and $\ell_L \leftrightarrow \ell_R$ in the diagrams.
Consequently, as can be seen from Fig. 1(a), the one-loop 13 element
of $M_D$ comes from the large tree-level 31 element. In the same
way, the 23 element gets a one-loop contribution that is an echo of
the large tree-level 32 element. (Loops involving gauge bosons that
are in ${\bf 1}$ or ${\bf 24}$ of $SU(5)$, do not change the $SU(5)$
representation of the fermions they couple to, and can be shown,
therefore, not to give non-zero contributions to the mass matrices
where there were zeros at tree level.) The diagram in Fig 1(a)
superficially looks divergent. However, the accidental symmetry
(discussed later) that makes $(M_D)_{13}$ vanish at tree level
guarantees that the loop is finite, as an exact calculation indeed
shows. The finiteness of this diagram is more obvious if we write it
in the form shown in Fig. 1(b). Neglecting higher order effects in
$\epsilon$, one finds that $(M_D)_{13} = (M_L)_{13} = \frac{9
g_U^2}{32 \pi^2} I(M_g^2/M_f^2) (M_D)_{31}$, where $g_U$ is the
gauge coupling at the unification scale, $I(x) \equiv \ln x/(x-1)$,
$M_g$ is the mass of the superheavy gauge boson in Fig. 1(a), and
$M_f$ is the mass of the heaviest fermion in the loop. (This
expression is valid in the limit of small mixing between the chiral
families ${\bf 16}_i$ and the extra fermion multiplets ${\bf 16} +
\overline{{\bf 16}} + {\bf 10}$.) The finite one-loop contributions
to $(M_D)_{23}$ and $(M_L)_{32}$ are larger by a factor $C_2/C_1$.
The other zero elements in the mass matrices get no contributions
from these one-gauge-boson-loop diagrams, because there is no
non-zero tree-level entry for them to ``echo".

\begin{picture}(360,180)
\thicklines
\put(60,60){\vector(1,0){30}}
\put(90,60){\line(1,0){30}}
\put(120,60){\vector(1,0){30}}
\put(150,60){\line(1,0){60}}
\put(240,60){\vector(-1,0){30}}
\put(240,60){\line(1,0){30}}
\put(300,60){\vector(-1,0){30}}
\put(180,60){\oval(120,80)[t]}
\put(180,60){\circle*{20}}
\put(210,100){\vector(-1,0){30}}
\put(70,45){${\bf 10}({\bf 16}_1)$}
\put(130,45){$\overline{{\bf 5}}({\bf 16}_1)$}
\put(195,45){${\bf 10}({\bf 16}_3)$}
\put(250,45){$\overline{{\bf 5}}({\bf 16}_3)$}
\put(160,110){${\bf 10}({\bf 45}_g)$}
\put(160,20){{\bf Fig. 1(a)}}
\end{picture}

\begin{picture}(360,180)
\thicklines
\put(30,70){\vector(1,0){25}}
\put(55,70){\line(1,0){25}}
\put(80,70){\vector(1,0){25}}
\put(105,70){\line(1,0){50}}
\put(180,70){\vector(-1,0){25}}
\put(180,70){\vector(1,0){25}}
\put(205,70){\line(1,0){50}}
\put(305,70){\vector(-1,0){50}}
\put(330,70){\vector(-1,0){25}}
\put(180,70){\oval(200,120)[t]}
\put(180,70){\circle*{5}}
\put(205,130){\vector(-1,0){25}}
\put(40,55){${\bf 10}({\bf 16}_1)$}
\put(90,55){$\overline{{\bf 5}}({\bf 16}_1)$}
\put(140,55){${\bf 5}({\bf 10})$}
\put(190,55){$\overline{{\bf 5}}({\bf 10})$}
\put(240,55){${\bf 10}({\bf 16}_3)$}
\put(290,55){$\overline{{\bf 5}}({\bf 16}_3)$}
\put(160,140){${\bf 10}({\bf 45}_g)$}
\put(130,70){\line(0,-1){30}}
\put(230,70){\line(0,-1){30}}
\put(175,80){$M_{10}$}
\put(110,25){$\langle {\bf 1}({\bf 16}_{1H}) \rangle$}
\put(210,25){$\langle \overline{{\bf 5}}({\bf 16}'_H) \rangle$}
\put(160,0){{\bf Fig. 1(b)}}
\end{picture}

\vspace{0.2cm}

\noindent
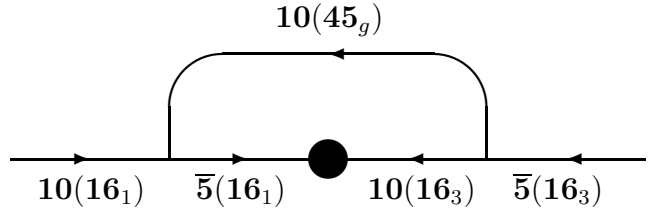
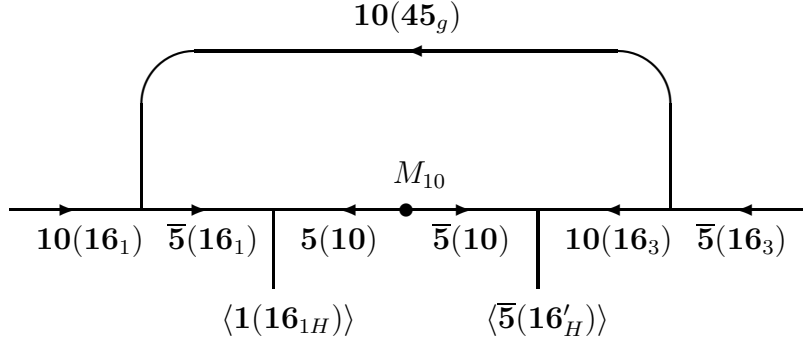
{\bf Figure 1.} The diagram in (a) shows how the 13 element of $M_D$ (or $M_L^T$)
arises radiatively from the tree-level 31 element, shown as the blob in the
center.  The ${\bf 10}({\bf 45}_g)$ in the loop is a superheavy gauge boson.
The diagram in (b) is more detailed and shows why the loop is finite.

\vspace{0.5cm}

The non-zero 13 element of $M_D$ and 31 element of
$M_L$ induced by the one-gauge-boson-loop diagrams are obviously not
enough to make these matrices have non-zero determinant. There must
also be a non-zero 22 (or 21) element for $M_D$ with an equal
non-zero 22 (or 12) element for $M_L$. The gauge loops do not
produce these. They can arise, however, in a rather simple way from
the one-Higgs-boson-loop diagram shown in Fig. 2.

\begin{picture}(360,180)
\thicklines
\put(60,75){\vector(1,0){30}}
\put(90,75){\line(1,0){30}}
\put(120,75){\line(1,0){30}}
\put(180,75){\vector(-1,0){30}}
\put(180,75){\vector(1,0){30}}
\put(210,75){\line(1,0){30}}
\put(240,75){\line(1,0){30}}
\put(300,75){\vector(-1,0){30}}
\put(180,75){\oval(120,80)[t]}
\put(180,75){\vector(0,-1){15}}
\put(180,60){\line(0,-1){10}}
\put(180,115){\vector(-1,0){30}}
\put(180,115){\vector(1,0){30}}
\put(180,115){\circle*{5}}
\put(70,60){${\bf 10}({\bf 16}_2)$}
\put(130,60){${\bf 10}({\bf 16})$}
\put(195,60){${\bf 10}({\bf 16})$}
\put(250,60){$\overline{{\bf 5}}({\bf 16}_2)$}
\put(160,40){$\langle {\bf 5}({\bf 10}'_H) \rangle$}
\put(130,125){${\bf 5}({\bf 10}_H)$}
\put(190,125){$\overline{{\bf 5}}({\bf 10}_H)$}
\put(160,0){{\bf Fig. 2}}
\end{picture}

\vspace{0.2cm}

\noindent
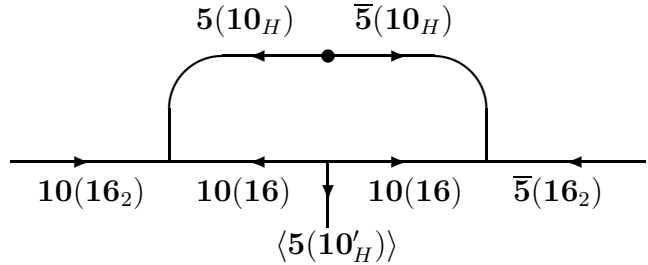
{\bf Figure 2.} A diagram showing how the 22 elements of the mass matrices can arise
radiatively through Higgs-boson loops.

\vspace{0.5cm}

Whereas the
gauge-loop diagrams discussed previously {\it must} exist, the
diagram in Fig. 2 only exists if certain couplings are present. In
particular, there must be the last term in Eq. (1), namely
$h ({\bf 16} \; {\bf 16}) \; {\bf
10}'_H$, and a Higgs-mass term of the form ${\bf 10}_H {\bf 10}_H$. The diagram in
Fig. 2 also gives a
22 element for the up-quark mass matrix $M_U$. However, if one assumes
that the $Y/2=+1/2$ and $Y/2= -1/2$ VEVs in $\langle {\bf 10}'_H
\rangle$ are of the same order, then the contribution of Fig. 2
to $m_c$ is only a few percent and thus negligible.

The full mass matrices, including the contributions from one-loop diagrams,
have the form

\begin{equation}
\begin{array}{ll}
M_U = \left( \begin{array}{ccc} 0 & 0 & 0 \\
0 & 0 & \epsilon/3 \\ 0 & - \epsilon/3 & 1 \end{array}
\right) \; m_U, \;\;\; & M_D = \left( \begin{array}{ccc}
0 & 0 & \delta_g \\ 0 & \delta_H & \epsilon/3 + \delta'_g \\
C_1 & C_2 - \frac{\epsilon}{3} & 1 \end{array} \right) \; m_D, \\ & \\
M_N = \left( \begin{array}{ccc} 0 & 0 & 0 \\
0 & 0 & - \epsilon \\ 0 & \epsilon & 1 \end{array}
\right) \; m_U, \;\;\; & M_L = \left( \begin{array}{ccc}
0 & 0 & C_1 \\ 0 & \delta_H & C_2 - \epsilon \\
\delta_g & \epsilon + \delta'_g & 1 \end{array} \right) \; m_D, \end{array}
\end{equation}

\noindent
where $\delta'_g = (C_2/C_1) \delta_g$. A fit of the first family masses and
mixings requires that $\delta_g$ and $\delta_H$ be approximately equal to
$10^{-2}$, which
is consistent with the magnitude of
one-loop effects. Note that $M_U$ still has rank = 2, so that the $u$ quark
is massless even at one-loop level. This is quite in accord with the fact that
$m_u/m_t \ll m_d/m_b$ and $m_e/m_{\mu}$. In order to reproduce the observed mass
of the $u$ quark, there has to be either a 11 element in $M_U/m_U$ that is $ \sim
3 \times 10^{-5}$
or 12 and 21 elements that are $\sim 4 \times 10^{-4}$. In either case, these
are
much smaller than one would expect from a typical one-loop diagram, but quite
consistent with a two-loop effect.

To see the pattern of masses and mixings for the first family, it is
convenient to make the same transformation of $M_D$ that eliminates the large
off-diagonal elements $C_1$ and $C_2$, and which at tree-level led to Eq. (7).
At one loop, it leads to

\begin{equation}
M'_D = \left( \begin{array}{ccc}
0 & -s_a \delta_g & c_a \delta_g \\ & & \\
-s_s \delta_H & -s_a (\epsilon/3 + \delta'_g) & c_a(\epsilon/3 + \delta'_g) \\
& +c_a c_s \delta_H & +s_a c_s \delta_H \\ & & \\
0 & 0 & 1/c_a \end{array} \right) m_D.
\end{equation}

\noindent
It is well-known that the empirical relation
$\tan \theta_C \simeq \sqrt{m_d/m_s}$ is obtained if the following
conditions are satisfied: $(M_U)_{12}, (M_U)_{21} \simeq 0$, $(M_D)_{11} \simeq 0$, and $|(M_D)_{12}| \simeq |(M_D)_{21}|$. In order to satisfy the last of these conditions
in this model, all that is required (as can be seen from Eq. (9)) is that $|\delta_H \sin \theta_{sol}| \simeq
|\delta_g \sin \theta_{atm}|$, which is natural, given that $\delta_g$ and $\delta_H$
are both one loop effects and that $\theta_{atm}$ and $\theta_{sol}$ are both
of order one.

The only other parameter that relates to the first family that remains to
be accounted for is $V_{ub}$. From Eq. (9), one sees that, because $|V_{ub}| \cong
|(M'_D)_{13}/m_b|$ and $|V_{us}| \cong |(M'_D)_{12}/m_s|$, one obtains the relation
$|V_{ub}| \cong |V_{us}| (m_s/m_b) \cot \theta_{atm}$. Using the facts that $m_s/m_b
\simeq 0.02$, $V_{us} \simeq 0.2$, and $\theta_{atm} \simeq \pi/4$, one obtains
$|V_{ub}| \simeq 0.004$, which is quite close to the measured value.

One can invert this equation to get a value for $\theta_{atm}$:
using the best fit values for $|V_{ub}|$ and $|V_{us}|$ and the
Georgi-Jarlskog value for $m_s/m_b$ (i.e. $m_{\mu}/3 m_{\tau}$), one
finds $\tan \theta_{atm} \cong 1.47$, corresponding to
$\theta_{atm} \simeq 56^{\circ}$. (Considering that there are
contributions to the atmospheric angle of $O(\epsilon)$ of about $10^{\circ}$
coming from the diagonalization of the neutrino mass matrix,
this value is reasonable.) Since $|V_{ub}| \cong \delta_g \cos^2
\theta_{atm}$ one obtains $\delta_g \cong 0.94 \times 10^{-2}$. From
the relation $\delta_H \sin \theta_{sol} \cong \delta_g \sin
\theta_{atm}$ and the measured value $\tan \theta_{sol} \cong 0.66$,
one then obtains $\delta_H \cong 1.4 \times 10^{-2}$.
The values of
$\theta_{atm}$ and $\theta_{sol}$ directly yield $C_1 \cong 0.8$ and
$C_2 \cong 1.23$.
If one substitutes $C_1 = 0.8$ into the expression for the loop diagram
in Fig 1(a), and uses for the unified gauge coupling $\alpha_U = 0.025$,
one finds $\delta_g \cong 0.72 \times 10^{-2} I(x)$. Since $I(x)$ is a slowly varying
function and $I(1) = 1$, this is nicely in accord with the value
$\delta_g = 0.94 \times 10^{-2}$ obtained by fitting the quark and lepton masses. Thus
the masses of the electron and $d$ quark are very well explained as radiative effects.

The value of $\epsilon$ is fixed to be 0.15 directly by
$m_c/m_t$ and the value of $m_U/m_D$ is fixed by $m_t/m_b$, so that all that
remains to be determined of the parameters appearing in Eq. (8) are the complex
phases.

Most of the complex phases can be rotated away from the matrices by
phase redefinitions of complete Standard Model multiplets, and so
have no low-energy implications. However, three phases cannot be
rotated away. These can be taken to be the phases of $\delta_H$,
$\delta'_g$, and the phase of the $\epsilon$ that appears with $C_2$
in the 32 element of $M_D$ and $M_L^T$. Call these phases
$\alpha_H$, $\alpha_g$, and $\alpha_{\epsilon}$. The last of these
has an effect only at subleading order in $\epsilon$ and so we will
ignore it. The phases $\alpha_H$ and $\alpha_g$, on the other hand,
enter importantly in the expressions for $m_s$ and $V_{cb}$. Fixing
these phases by these two measured quantities gives a definite
prediction for the CKM CP-violating phase $\delta_{CKM}$. A fit
gives roughly $\alpha_H \sim 0$, $\alpha_g \sim \pi/2$. The angle
$\phi_3$ in the unitarity triangle then comes out to be about
$32^{\circ}$, which is too small. It should be noted, however, that
the CP-violating phase is more sensitive to the values of the model
parameters than most of the other observables. Moreover, two-loop
effects can significantly affect it. For example, the type of
two-loop effects needed to generate the mass of the $u$ quark could
well generate a 12 element for $M_U$ of order $10^{-4}$. That would
have an effect on $V_{us}$, and thus on the best-fit value of
$\delta_g$, of order $20 \%$. That would in turn affect the
CP-violating phase (which comes predominantly from the phase of
$\delta'_g$) by a factor of the same order.

The foregoing numbers are based on a rough fit, and a more careful
analysis must be done. This would require (a) doing the
renormalization group running of the mass and mixing angles from the
unification scale down to low scales within a breaking scheme for
$SO(10)$ that reproduces the unification of gauge couplings, and (b)
doing a global fit to all the masses and mixings. In the limit we
are taking (of small mixing between the chiral families ${\bf 16}_i$
and the real extra fermions ${\bf 16} + \overline{{\bf 16}} + {\bf
10}$), there are 9 dimensionless model parameters: $m_U/m_D$, $C_1$,
$C_2$, $\epsilon$, $\delta_g$, $\delta_H$, and the phases
$\alpha_g$, $\alpha_H$, and $\alpha_{\epsilon}$, the last being
relatively unimportant. These must fit 14 dimensonless observables:
7 mass ratios of the charged fermions (not counting the $u$ quark,
which is still massless at one-loop level), and 7 mixing parameters.

Finally, something must be said about the structure of the Yukawa sector given in
Eq. (1). It is easily seen that the terms given in Eq. (1) leave accidental global
abelian symmetries (some of which may be gauged) that prevent other Yukawa terms that
could be dangerous.  It is these accidental symmetries that make the zeros in the
mass matrices in Eq. (5) ``technically natural", and guarantee that the loop effects
$\delta_g$, $\delta'_g$ and $\delta_H$ are ``finite and calculable".

In conclusion, we have shown that it is possible to construct a rather simple
and predictive model of quark and lepton masses based on the idea of a ``radiative
hierarchy". All the dimensionless parameters in this model take values
of order one, except $m_U/m_D$. That ratio determines the overall scale
of $Y/2 = +1/2$ masses relative to $Y/2 = -1/2$ masses (i.e. ``up" to ``down"),
and may have an ``anthropic" or ``landscape" explanation \cite{bk}. The large
interfamily hierarchies are achieved without ad hoc tuning of parameters to
small values. A striking aspect of the model is the way that the ``lopsided"
structure (i.e. $C_1 \sim C_2 \sim 1$)
explains explains so many features of the light fermion spectrum.
In particular, it explains (a) the large solar and atmospheric mixing angles and the
small value of $\theta_{13}$ (the so-called ``bilarge" pattern of neutrino mixing),
(b) the emergence
of the Georgi-Jarksog factor of 3 (which would be $\cong 9$ if it were not for the
fact that $C_2 \gg \epsilon$), (c) the largeness of $m_s/m_b$ and
$m_{\mu}/m_{\tau}$ compared to $m_c/m_t$, (d) the fact that $V_{cb}$ is of order
$m_s/m_b$ rather than $\sqrt{m_s/m_b}$, as it is in models with symmetric
mass matrices, (e) the fact that $V_{ub}$ comes out with the correct magnitude
($\simeq V_{us} (m_s/m_b) \simeq 0.004$), and (f) the fact that $m_u/m_t
\ll m_d/m_b, m_e/m_{\tau}$, which is a consequence of the fact that the lopsided
entries do not appear in $M_U$, so that the one-gauge loop diagrams cannot change
its rank, and the $u$ quark must obtain mass from two-loop order. In other words many
of the peculiar features of the spectrum are traceable to a single simple
feature of the mass matrices.

\end{document}